\documentclass[12pt]{article}
\usepackage{latexsym} 

\newcommand{\beq}{\begin{equation}}
\newcommand{\eeq}[1]{\label{#1}\end{equation}}
\newcommand{\bea}{\begin{eqnarray}}
\newcommand{\eea}[1]{\label{#1}\end{eqnarray}}
\begin{document}

\markboth{M. Porrati}{Higgs phenomenon for the graviton in AdS space}

%
%

\title{{HIGGS PHENOMENON FOR THE GRAVITON IN ADS SPACE}}
\author{M. PORRATI\\ Department of Physics, New York University\\ 4 Washington Pl. New York, NY 10003, USA}



\maketitle


\begin{abstract}
In this review, we summarize recent findings that show how standard 4-d
Einstein gravity coupled to a conformal field theory can become massive in 
Anti de Sitter Space. Key ingredients in this phenomenon are non-standard 
``transparent'' boundary condition given to the CFT fields and the fact that
AdS space is not globally hyperbolic, due to the presence of a time-like 
boundary.
\end{abstract}

\section{Ward Identities in AdS}	
A widely held misconception about gauge theories in general, and generally
covariant ones in
particular, is that gauge invariance forbids an explicit mass term for the
gauge field. In the case of gravity, the gauge field is the graviton, and
invariance under general coordinate transformations translates into a set of 
Ward identities. Actually, these identities do not depend on the fact that 
gravity is dynamical; they also hold for a field theory on an arbitrary {\em
fixed} space-time background. In case the field theory possesses additional 
symmetries, additional Ward identities will hold.

Let us be specific and consider a 4-d conformal field theory on an arbitrary
background with metric $g_{mn}$. 
Let us denote by $W[g_{mn}]$ the generating functional of the connected Green
functions of the stress-energy tensor. To wit:
\begin{eqnarray}
{\delta W \over \delta g_{mn}}&=& \langle 0|T^{mn}|0 \rangle, \nonumber \\
{\delta^2 W \over \delta g_{mn}\delta g_{pq}} &=& \langle 0| 
T^{mn} T^{pq}| 0\rangle
-\langle 0| T^{mn} |0 \rangle \langle 0|T^{pq}|0 \rangle\equiv \Sigma^{mn,pq}, 
\label{1}
\end{eqnarray}
and so on.

Let us expand $W$ around a background $\bar{g}_{mn}$ 
($h_{mn}\equiv g_{mn}-\bar{g}_{mn}$):
\begin{eqnarray}
W[\bar{g}_{mn} + h_{mn}]&=& W[\bar{g}_{mn}] + \int d^4x \sqrt{-\bar{g}}
\langle 0| T^{mn} | 0 \rangle h_{mn} + \nonumber \\
&& +{1\over 2}\int d^4x \sqrt{-\bar{g}} \int d^4y \sqrt{-\bar{g}} h_{mn}(x)
\Sigma^{mn,pq}(x,y) h_{pq}(y) +... 
\label{2}
\end{eqnarray}
It is important to recall that $W[g]$ is only defined up to local terms in 
$g_{mn}$; equivalently, the Green functions are determined only up to
local contact terms.

General covariance and conformal invariance translate into two Ward Identities:
\begin{equation}
g_{mn}{\delta W \over \delta g_{mn}}= A[g], \qquad
D_m {\delta W \over \delta g_{mn}}=0.
\label{3}
\end{equation}
Here $A[g]$ is the conformal anomaly, whose specific form will be given later
to the necessary extent. 
$D_m$ is the standard Riemann covariant derivative. 
By expanding around $\bar{g}_{mn}$ up to quadratic order in the fluctuation
$h_{mn}$, and by denoting with $\bar{D}_m$ the background
covariant derivative, we find
\begin{equation}
\bar{g}_{mn}\Sigma^{mn,pq}(x,y)=0, \qquad
{\bar D}_m\Sigma^{mn,pq}(x,y)=0, \qquad x\neq y.
\label{4}
\end{equation}

We can do better than that. Indeed, by a judicious choice of {\em finite, 
local} counter-terms, we can completely cancel the contact terms when the
background is any of the three Lorentzian-signature, 
maximally symmetric spaces in 4-d: de Sitter (dS), Minkowsky (M) or Anti de 
Sitter (AdS)~\cite{p}.

First of all, we notice that maximal symmetry of the background implies that
the VEV of the stress energy tensor is proportional to the background metric,
$\langle 0| T_{mn} | 0 \rangle = C \bar{g}_{mn}$. Furthermore, we can use the 
ambiguity in the definition of $W[g]$ and define a new generating functional
\begin{equation}
W'[g]=W[g] + 2C\int d^4 x \sqrt{-g}.
\label{5}
\end{equation}
This new functional does not contain any term linear in the fluctuations around
the background, i.e. $W'[g]=W'[\bar{g}] + O(h^2)$. The Ward identity due to
general covariance, Eq.~(\ref{3}), now implies that $\Sigma^{mn,pq}$ is 
exactly transverse, without any contact term. 
\begin{equation}
{\bar D}_m\Sigma'^{mn,pq}(x,y) =0 \mbox{ everywhere}, \qquad
\Sigma'^{mn,pq}\equiv {\delta^2 W' \over \delta g_{mn}\delta g_{pq}}.
\label{6}
\end{equation}
The conformal Ward identity for $W'[g]$ is
\begin{equation}
g_{mn}{\delta W' \over \delta g_{mn}}= A[g]-A[\bar{g}]=\alpha C_{abcd}C^{abcd}
+\beta (R_{ab}-{1\over 4}g_{ab}R)^2+
\gamma(R^2-\bar{R}^2),
\label{7}
\end{equation}
with $\alpha,\beta,\gamma$ known constants. 
Expanding $W'[g]$ to linear order in $h_{mn}$, we find
\begin{equation}
\bar{g}_{mn}\Sigma'^{mn,pq} * h_{pq}=2\gamma \bar{R}R^L(h)+ O(h^2).
\label{8}
\end{equation}
$R^L(h)$ is the linearized scalar curvature and $*$ is the convolution 
in the 4-d coordinates $x^m$. We can finally define a 
generating functional $W''[g]$ by
\begin{equation}   
W''[g]= W'[g]+ 2\gamma \bar{R}\int d^4 x \sqrt{-g}\left(R-{1\over 2}\bar{R}
\right).
\label{9}
\end{equation}
The conformal Ward identity on $W''[g]$, expanded to linear order in $h_{mn}$
implies
\begin{equation}
\bar{g}_{mn}\Sigma''^{mn,pq}=0 \qquad \mbox{everywhere}.
\label{10}
\end{equation}
We can summarize what we did so far as follows: in a CFT on a maximally 
symmetric space, we can make the 2-point 
function of the stress-energy tensor transverse-traceless everywhere, including
at contact points, by adding finite contact terms. These terms explicitly
break conformal invariance. Notice that these terms are the standard Einstein
action and the cosmological constant term. So, when we will let the metric 
$g_{mn}$ fluctuate dynamically, they will generate finite renormalizations
of the cosmological and Newton constants.

\section{The Graviton Mass: General Properties}
As anticipated at the end of the previous section, we want to couple our 
(cutoff) CFT to gravity. We will treat gravity perturbatively, by discarding 
graviton loops, but we will include all matter loops. This approximation is
consistent whenever the cutoff $\Lambda$ is well below the Planck scale 
($\Lambda\ll M_{Pl}$).

The result of integrating out the CFT fields is the effective action
\begin{equation}
\Gamma_{eff}[g]=16\pi G_B\int d^4x \sqrt{-g} (R-2\Lambda_B) + W[g].
\label{11}
\end{equation}  
$W[g]$ is the result of integrating out the CFT degrees of freedom, so it
is precisely the generating functional of the connected Green functions of
the stress-energy tensor, as in the previous section.
Since $W[g]$ is defined only up to contact terms, the splitting of 
$\Gamma_{eff}[g]$ in between the ``bare'' Einstein action and $W[g]$ is 
ambiguous. We can fix this ambiguity by defining
\begin{equation}
\Gamma_{eff}[g]=16\pi G\int d^4x \sqrt{-g} (R-2\Lambda) + W''[g].
\label{12}
\end{equation}
With this choice, $G$ is the physical Newton constant, and $\Lambda$ is
the physical cosmological constant. By this we mean in particular that 
$\bar{R}_{ab}=\Lambda \bar{g}_{ab}$ solves the equations of motion derived 
from the action $\Gamma_{eff}[g]$. As before we denote by $\bar{g}_{mn}$ 
the metric of a maximally symmetric space.

By expanding the action to quadratic order around $\bar{g}_{mn}$, we get the
linearized equations of motion for the fluctuation $h_{mn}$. In particular,
the transverse-traceless part of the fluctuation, $h^{TT}_{mn}$, obeys
\begin{equation}
(\Delta -2 \Lambda) h_{mn}^{TT} + 32\pi G {\Sigma''}_{mn}^{pq}*h^{TT}_{pq}=0.
\label{13}
\end{equation}
Here, $\Delta$ denotes the Lichnerowicz operator~\cite{l}, 
\begin{equation}
\Delta h_{mn}=-\Box h_{mn} -2R_{mrnp}h^{rp} +
2R_{(m}^r h_{n)r}.
\label{14}
\end{equation}
In AdS space, a free spin-2 field $\phi^M_{mn}$ of mass $M$ obeys the equation
of motion
\begin{equation}
(\Delta -2 \Lambda) \phi_{mn}^{M\,TT}= -M^2 \phi_{mn}^{M\,TT}.
\label{15}
\end{equation}
Also, $\Sigma''^{mn,pq}$ is the two-point function of the stress-energy tensor
on the background metric, so it commutes with $\Delta$ and can be diagonalized
simultaneously with it~\footnote{To prove this,
use a spectral representation obtained by inserting a complete set of 
intermediate states: $\Sigma''^{mn,pq}(x,y)=\sum_E\langle 0 | T^{mn}(x)
|E\rangle\langle E| T^{pq}(y)|0\rangle$. In the case of interest here, AdS 
space, this basis will be described in more details in the next section.}.

If we combine Eqs.~(\ref{14},\ref{15}), and we define
${\Sigma''}_{mn}^{pq}*\phi_{pq}^{M\,TT}\equiv \Sigma(M^2)\phi_{pq}^{M\,TT}$,
we find that $h_{mn}$ describes spin-2
excitations with mass given by the solution of the equation
\begin{equation}
-M^2 + 32\pi G \Sigma(M^2)=0.
\label{16}
\end{equation}
This equation says that a massless graviton exists when $\Sigma(M^2)=0$.
On the other hand, when $\Sigma(M^2)$ is continuous around zero and
$\Sigma(0)> 0$, Einstein gravity coupled to a CFT does not possess any
massless graviton. The mass of the lightest spin-2 excitation is instead
\begin{equation}
M_{min}\approx \sqrt{32\pi G\Sigma(0)}.
\label{17}
\end{equation}
This is at best a {\em very} small mass! To appreciate this fact, consider 
that $\Sigma(0)$ is independent of $G$: it is a correlator in a CFT on a 
{\em fixed} background with cosmological constant $\Lambda$. 
Purely by dimensional analysis, we have $\Sigma(0)=O(\Lambda^2)$, since
$\Lambda$ is the only scale of the CFT. Equation~(\ref{17}) then gives the
following estimate for the mass of the lightest spin-2 excitation
\begin{equation}
M_{min}=O\left(\sqrt{G\Lambda^2}\right)=O\left(|\Lambda|
\over M_{Pl}\right) \ll \sqrt{|\Lambda|}.
\label{18}
\end{equation}
The last inequality holds whenever the radius of curvature of the space,
$L\equiv \sqrt{3/|\Lambda|}$ is much larger than the Planck length 
$L_{Pl}\equiv M_{Pl}^{-1}$.

Up to now, we have not proven that $\Sigma(0)\neq 0$; indeed, nothing we said 
so far is specific to AdS space. To prove that the graviton does indeed 
get a mass we need to study more carefully the properties of (free) fields 
in AdS space. This is the subject of next section.       

\section{Free Particles in AdS Space}
To proceed further we need to review some facts about 
positive-energy representations of the $AdS_4$ isometry group, $SO(2,3)$.

These representations were classified in~\cite{e} (see also~\cite{n} for a 
clear review). In the decomposition $SO(2,3)\rightarrow SO(2)\times SO(3)$, 
the generator of $SO(2)$ is the $AdS_4$ energy, while angular momentum is 
given by the generators of $SO(3)$. A unitary, irreducible, positive-weight 
representation of $SO(2,3)$ (UIR), $D(E,s)$, is labeled by the energy $E$ 
and spin $s$ of its (unique, up to spin degeneracy) lowest-energy 
state~\footnote{$E$ is the dimensionless energy, measured in units of the 
$AdS$ curvature radius $L$.}. 
Free fields form irreducible representations of $SO(2,3)$. A
conformal scalar can belong to either the $D(1,0)$ or the $D(2,0)$. 
A conformal (massless) spin-1/2 fermion belongs to a $D(3/2,1/2)^{\pm}$,
while a massless vector (also conformal) belongs to a 
$D(2,1)^{\pm}$~\cite{f,bf}.
The label $\pm$ denotes the parity of the UIR.

Massless representations of spin $s> 0$ 
have $E=s+1$~\cite{f,bf,n}. Massive unitary 
representations of spin larger than zero have $E>s+1$. In the limit
$E\rightarrow s+1$, the UIR $D(E,s)$, $s\geq 1$ 
becomes reducible~\cite{e,f}:
\begin{equation}
D(E,s) \rightarrow D(s+1,s) \oplus D(s+2,s-1), \qquad E\rightarrow s+1.
\label{19}
\end{equation}

Eq.~(\ref{19}) encodes the group theoretical aspect of the Higgs phenomenon in
$AdS_4$: when a spin-$s$ field, $s\geq 1$, becomes massive, it ``eats''
a spin-$(s-1)$ boson. Notice than for $s=0$ this boson is in a $D(3,0)$, i.e. 
it is a minimally-coupled scalar~\cite{bf}. For spin 2, it is a {\em massive}
vector in the $D(4,1)$~\footnote{Recall that in Minkowsky space, instead, the
Higgs phenomenon for a spin 2 requires a massless vector and a massless 
scalar. They together provide 3 degrees of freedom, as it does our
massive vector in $AdS_4$.}. 

Now, this $E=4$, spin-1 field is the Goldstone field for the graviton, so it
must have a nonzero matrix element with the state $T_{mn}|0\rangle$. 
Equivalently, the connected two-point function of the graviton must have a 
pole corresponding to a vector belonging to $D(4,1)$.
The wave equation obeyed by such a vector is
$(\Delta -2\Lambda)A_m =0$~\cite{p}. So, a Goldstone vector exists when
\begin{equation}
  {\Sigma''}_{mn}^{pq} * h_{pq} = {2C\over \Delta -2\Lambda} 
D_{(m}D^l h_{n)l} + .....,
\label{20}
\end{equation} 
and $C$ is a nonzero constant, that determines the strength of the 
Goldstone boson coupling.
Since ${\Sigma''}_{mn}^{pq} * h_{pq}$ is transverse and traceless, it is 
proportional to the projector over TT states, $\Pi_{mn}^{pq}$. This projector 
was computed, for instance, in~\cite{p}. Its relevant part reads
\begin{equation}
\Pi_{mn}^{pq}* h_{pq}= h_{mn} + {2\over \Delta -2\Lambda} 
D_{(m}D^l h_{n)l}  +.....\; .
\label{21}
\end{equation}
In the previous section, we saw that 
$\Sigma''^{mn,pq}=\Sigma(\Delta-2\Lambda)\Pi^{mn,pq}$, so
Eqs.~(\ref{20},\ref{21}) and the continuity of $\Sigma(M^2)$ near $M^2=0$
imply $C\approx \Sigma(0)$.

Not surprisingly, we get the same condition that we found at the end of
Section 2: a Goldstone vector exists --i.e. the graviton acquires a mass--
iff $\Sigma(0)\neq 0$. 

The projector $\Pi_{mn}^{pq}$ has also an additional, spin-zero pole at 
$\Delta=4\Lambda/3$~\cite{dls}. 
To see whether this scalar is a physical degree of freedom or a ghost, 
it is useful to introduce an external source with 
(covariantly conserved) stress-energy 
tensor $T_{mn}$, and use the fact that action $\Gamma_{eff}$ in Eq.~(\ref{11})
is invariant under general coordinate transformations to decompose the metric
fluctuations as 
$h_{mn}=h_{mn}^{TT} + \bar{g}_{mn} \psi $. In this gauge,
the equations of motion derived from $\Gamma_{eff}$ in the presence of the
external source are
\begin{eqnarray}
(\Delta -2 \Lambda) h_{mn}^{TT} + 32\pi G \Sigma(\Delta-2\Lambda)* h^{TT}_{pq}
&=&2\Pi_{mn}^{pq}*T_{pq}, \nonumber \\ 
(3\Delta    -4\Lambda) \psi &=& T.
\label{22}
\end{eqnarray}
On the conserved source $T_{mn}$ we also have
\beq
\Pi_{mn}^{pq}*T_{pq}= T_{mn} +{1\over 3} \bar{g}_{mn} T +(
D_m D_n + \bar{g}_{mn} \Lambda/3) (3\Delta - 4\Lambda)^{-1}T.
\label{22'}
\end{equation}
The one-graviton 
interaction between two conserved sources, $T_{mn}$, $T'_{mn}$ is given by
the amplitude $A=(1/2)T'^{mn}h_{mn}(T)$,
which has indeed a scalar pole at $\Delta=4\Lambda/3$:
\begin{equation}
A= T'^{mn}h_{mn}\approx 8\pi G {\Sigma(-2\Lambda/3)\over\Lambda} T' 
(\Delta-4\Lambda/3)^{-1}T.
\label{23}
\end{equation}
The residue is always small, $O(L_{Pl}^2/L^2)$. It is physical when
$\Sigma(-2\Lambda/3)\leq 0$. Since positivity of the graviton residue
requires instead $\Sigma(0)\geq 0$, we conclude that $\Sigma(M^2)$ cannot be
constant in the range $0\leq M^2 \leq -2\Lambda/3$ in a unitary theory. 
This is possible in our case, since the CFT 
coupled to gravity provides us with an infinite number of degrees of freedom
with mass $O(\sqrt{|\Lambda|})$.

Let us come back to the search for the Goldstone vector, in the special case
where the CFT is free (before coupling it to gravity, i.e. in the limit $M_{Pl}
\rightarrow \infty$).
In a free conformal field theory, $T_{mn}$ is 
quadratic in the fields, so the Goldstone vector is a bound state in 
the product of two free fields~\footnote{In Minkowsky space this is obviously 
absurd since non-interacting two-particle states form a continuum. 
In Anti de Sitter space, instead, free particles do form bound states, since 
the AdS energy spectrum is discrete.}.
A necessary condition for the existence of a Goldstone boson is then that 
the $D(4,1)$ UIR appears in the tensor 
product of the UIRs to which the CFT fields belong:
\begin{equation}
D(4,1)\subset D(E,s)\otimes D(E,s).
\label{24}
\end{equation}
 
Let us examine separately free conformally-coupled fields of spin 1,
1/2, and 0.
\begin{description}
\item{spin 1}
Massless spin-1 fields are conformal; they belong to the $D(2,1)$~\cite{f,bf}.
The tensor product of $SO(2,3)$ UIRs was found by Heidenreich 
in~\cite{h}. For $D(2,1)$, he found
\bea
D(2,1)\otimes D'(2,1)&=& \sum_{n=0}^\infty D(4+n,0) \oplus 
\sum_{n=0}^\infty D(4+n,1) \oplus  \nonumber \\
&& \sum_{S=0}^\infty \bigg[D(4+S,2+S)
\oplus \sum_{n=0}^\infty 2D(5+S+n,2+S)\bigg].
\eea{25}
In our case, since we are tensoring two identical bosons, some of the
representations that appear in the tensor product above are absent.
For instance, the ground state of the
$D(4,1)$ that appears in the tensor product above is antisymmetric in its 
arguments ($\sim \epsilon_{ijk}A^iA^j$), so it is forbidden by Bose 
statistics. This means that the entire $D(4,1)$ is absent.
\item{spin 1/2}
The massless (conformal) spin-1/2 field belongs to the $D(3/2,1/2)$.
Tensoring two different $D(3/2,1/2)$, Heidenreich finds~\cite{h}
\bea
D(3/2,1/2)\otimes D'(3/2,1/2) &=&  \sum_{n=0}^\infty D(3+n,0) \oplus
\sum_{S=0}^\infty \bigg[D(3+S,1+S)\oplus \nonumber \\ &&
\sum_{n=0}^\infty 2D(4+S+n,1+S)\bigg].
\eea{26}
In this tensor product, the $D(4,1)$ appears twice. By taking into account
Fermi statistics when tensoring two identical representations, we get rid of 
one of them. The other one cannot appear in the stress-energy tensor since it 
has the wrong parity~\cite{p}. 
To arrive at this result we first notice that the 
stress-energy tensor of a free CFT made of 
several fields of spin $s\leq 1$ is given by the sum 
$T_{mn}= \sum_iT^i_{mn}$.  The $i$-component of this sum is the
stress-energy tensor of either a real vector, a real scalar, or a Majorana
fermion. To preserve the Majorana condition ($\psi=C\psi^*$, $C=$ charge 
conjugation), the field $\psi(x)$ must transforms as follows under parity: 
$\psi(t,{\bf x})\rightarrow \eta \gamma^0 \psi(t,-{\bf x})$, $\eta=\pm 1 $.
The fermion field $\psi$ can be expanded in spherical 
waves. Its positive-frequency part (with respect to the 
global AdS time $t$) is~\cite{bf}
\beq
\psi_{pf} = \sum_{k=0}^\infty\sum_{n=0}^\infty e^{-i\omega t/L}
a_{\omega j m}
\chi_{\omega j m}^\pm,\qquad j=1/2+k,\qquad \omega=1 +j+n. 
\eeq{27}
The operators $a^\dagger_{\omega j m}, a_{\omega j m}$ 
respectively create and annihilate states  
of definite energy $\omega/L$ and angular momentum $j$, belonging to the
$D(3/2,1/2)^{\pm}$.
The superscript $\pm$ labels the parity of the representation. 
More precisely,  when $\psi$ belongs to the $D(3/2,1/2)^+$, the 
spherical waves $\chi_{\omega j m}^+$ in Eq.~(\ref{27}) 
transforms under parity as
$\chi_{\omega j m}^+\rightarrow i(-)^{\omega-3/2}\chi_{\omega j m}^+$. 
Analogously, for $D(3/2,1/2)^-$,   
$\chi_{\omega j m}^-\rightarrow -i(-)^{\omega-3/2}\chi_{\omega j m}^-$.

The parity a UIR 
is fixed by the parity of its ground state. The assignments given
above show immediately that the parity of the ground state of the
$D(4,1)$ in the tensor product of either $D(3/2,1/2)^+ \otimes D(3/2,1/2)^+$ 
or $D(3/2,1/2)^- \otimes D(3/2,1/2)^-$ is +1. This is the parity of a 
{\em pseudo-vector}, while the $D(4,1)$ contained in $T_{mn}$ must be a 
true vector, with parity $-1$.  This can be seen most easily by noticing that
$T_{mn}$ is a true tensor and that 
the $D(4,1)$ we are after must appear in it as follows
\beq
T_{mn}=D_{(m}A_{n)} +.... ,\qquad (\Delta -2\Lambda)A_m=0.
\eeq{28}
Equivalently, we may notice that with a single Majorana fermion we cannot 
form a vector, as $\bar{\psi}\gamma^m\psi=0$, but we can form the 
pseudo-vector $\bar{\psi}\gamma^m\gamma^5\psi$.

\item{spin 0}
Scalars belong to $D(E,0)$, $E\geq 1/2$.
The tensor product of two spin zero representations of $SO(2,3)$ is~\cite{h}
\beq
D(E_1,0)\otimes D(E_2,0)= \sum_{S=0}^\infty \sum_{n=0}^\infty
D(E_1+E_2+S+2n,S).
\eeq{29}
Here $E_1,E_2>1/2$. When $E=1/2$, the representation degenerates, becoming a 
singleton~\cite{ff,d}, namely a representation that propagates only boundary
degrees of freedom and cannot be represented as a standard 
local field living in the bulk of $AdS_4$. We will not consider it further.
When $E_1=E_2=E>1/2$, a $D(4,1)$ exists in the tensor product 
$D(E,0)\otimes D'(E,0)$ only for $E=3/2$. If the two representations are
identical, $D(4,1)$ is eliminated by Bose statistics [its would be ground 
state is in reality a descendant belonging to the $D(3,0)$].
\end{description}
This is not the end of the story though, 
since the $D(4,1)$ appears in the tensor
product of two representations of different energy. In particular, 
it appears in the product $D(1,0)\otimes D(2,0)$.
A conformally-coupled scalar field belongs entirely 
either to the $D(1,0)$ or the 
$D(2,0)$ when it is given reflecting boundary conditions at the boundary of
$AdS_4$~\cite{bf}. 
This is not the most general condition one can give. In particular,
one can give so-called ``transparent'' boundary conditions. They are obtained
by noticing that $AdS_4$ is conformal to half the Einstein static universe.
So, one can obtain an (over-complete) set of solutions for the conformal 
scalar wave equation by solving it in the static Einstein universe, and then
conformally transforming the solution to AdS. These boundary conditions were
studied first in~\cite{ais}. Physically, these conditions correspond to the 
possibility of letting the boundary exchange energy and momentum with the 
interior of the AdS space. This could be achieved by introducing 
a 3-d defect CFT localized  
at the boundary. This possibility may seem bizarre, but it is actually 
required~\cite{br} in order 
to give a holographic interpretation to certain warped
compactifications of 5-d gravity~\cite{kr,kr2,p2,dwfo,br,awfk}. 

\section{The Graviton Mass: Explicit Computation}
Transparent boundary conditions imply that a conformal scalar can belong to
the reducible representation $D(1,0)\oplus D(2,0)$. This changes the form
of its two-point function. 

The 2-point function of a conformally coupled scalar field 
is best written by representing
$AdS_4$ as the cover of a hyperboloid in a  5-d space with signature $(2,3)$:
\beq
X^M X^N \eta_{MN} =-L^2, \qquad \eta=\mbox{DIAG}(-1,-1,1,1,1).
\eeq{30}
Here, $X^M$ is a homogeneous coordinate. The distance between two points on
the hyperboloid, with coordinates $X^M$ and $Y^M$,  
is a function of $Z\equiv X^MY_M /L^2$ only. The scalar propagator, for 
arbitrary boundary conditions, reads
\beq
\Delta(Z)={1\over 4\pi^2L^2}
\left(\alpha {1\over Z^2-1} + \beta {Z\over Z^2-1}\right). 
\eeq{31}
Standard, reflecting boundary conditions are $\alpha=0, \beta=1$ [$D(1,0)$]
or $\alpha=1, \beta=0$ [$D(2,0)$]. ``Transparent'' boundary conditions are 
instead $\alpha=\beta=1/2$~\cite{p}.

Once the two-point function of the scalar is known, it is  relatively 
straightforward to find the two-point function of its (improved) stress-energy
tensor, and from this last quantity the mass of the graviton.
The most subtle point in the computation 
is that, in order to determine the mass unambiguously, one must 
look for the non-local term [cfr. Eq.~(\ref{20})]
\beq
\langle 0| T_{mn} (0) T_{pq}(x)|0 \rangle = {1 \over 2}
(0|{C\over \Delta -2\Lambda} \bar{g}_{mp} D_n D_q |x) + 
(m\leftrightarrow n, p\leftrightarrow q ) +...
\eeq{32}

In fact, it is only by looking at a non-local term that one can distinguish 
unambiguously between a true
mass term and a spurious term that can canceled by redefining the 
contact terms in the two-point function of the stress-energy tensor. 

In homogeneous coordinates, at large distances ($Z\rightarrow \infty$), 
the right-hand side of 
Eq.~(\ref{32}) goes as $Z^{-4}+ O(Z^{-6})$. The coefficient of
the $Z^{-4}$ term is $\Sigma(0)$, from which we can get the graviton mass
thanks to Eq.~(\ref{16}). The computation of ref.~\cite{p} gives
\beq
\Sigma(0)= {K\over L^4} \alpha\beta.
\eeq{33} 

The nonzero numerical constant $K$ is independent of $L,\alpha,\beta$.
Eq.~(\ref{33}) manifestly shows that a nonzero graviton mass arises only when
the matter CFT is given non-standard, non-reflecting boundary 
conditions~\footnote{The graviton has instead standard boundary conditions!}.

Ref.~\cite{dls} extended the computation of $K$ to the case of a free $N=4$ 
super Yang-Mills theory. The numerical value for $K$ found there is the same
that one obtains when the $N=4$ theory is strongly coupled, 
in which case the computation can be done using the
holographic duality~\cite{p2} with the Karch-Randall 
compactification~\cite{kr,kr2,m}. The reason behind this equality is quite
mysterious, since it is not due to any known non-renormalization theorem. 
At present, indeed, it is not clear if this equality persists when the 
gauge coupling of the $N=4$ super Yang-Mills theory is finite and nonzero.

\section{Conclusions}
In this review, we re-examined the possibility of giving a mass to the graviton
in Anti de Sitter space. 
We pointed out that Ward identities do not forbid a 
graviton mass. We then proceeded to examine
the conditions that allow a gravitational Higgs mechanism in AdS, 
and we gave a model-independent estimate of the graviton mass induced by a
CFT.

Next, we recalled  
that unlike Minkowsky space, $AdS_4$ allows even a free theory to form bound 
states, owing to the discreteness of the AdS energy spectrum. We investigated
free CFTs with spin not greater than 1. We found that, in order
to have a Goldstone particle in the stress-energy tensor, we had to impose
non-standard (i.e. non-reflecting) boundary conditions on the fields of our
free CFT. Similar boundary conditions are not only allowed, 
but indeed necessary, to interpret holographically the KR 
model~\cite{kr,p2,kr2,br,awfk}. 

We considered next a free conformal scalar in 
$AdS_4$, and we found 
that, even in that very simple example, nonstandard boundary conditions do 
produce a Goldstone boson that gives a mass to the graviton, when the CFT is 
coupled to standard Einstein's gravity. 

In~\cite{kr,p2}, it was shown that the same 
phenomenon happens when gravity is coupled to a {\em strongly} interacting CFT.
In both cases the key ingredient is the boundary conditions imposed on the CFT.
They must allow energy and momentum to flow in and out of $AdS_4$ through its
boundary.
\section*{Acknowledgments}
The author 
is supported in part by NSF through grants PHY-0070787 and PHY-0245068.


\end{document}